\documentclass[iop,revtex4]{emulateapj}

\usepackage{apjfonts}
\usepackage{times}

\usepackage{amsmath}
\usepackage{graphicx}
\usepackage{natbib}
\usepackage{epstopdf} 
\usepackage{subfigure}
\usepackage{color}
\usepackage{lscape}
\usepackage{helvet}

\shorttitle{}
\shortauthors{Shin et al.}

\newcommand{\ergs}{erg~s$^{\rm -1}$}
       
\newcommand{\angstrom}{\textup{\AA}}
\begin{document}
\title{A catalog of X-ray point sources in the Abell 133 region}
\author{Jaejin Shin$^{1}$, Richard. M. Plotkin$^{2,3}$, Jong-Hak Woo$^{1,5}$, Elena Gallo$^{3}$, and John S. Mulchaey$^{4}$}

\affil{
$^1$Astronomy Program, Department of Physics and Astronomy, 
Seoul National University, Seoul, 151-742, Republic of Korea\\
$^2$International Centre for Radio Astronomy Research{--}Curtin University, GPO Box U1987, Perth, WA 6845, Australia\\
$^3$Department of Astronomy, University of Michigan, 1085 South University Avenue, Ann Arbor, MI 48109, USA\\
$^4$The Observatories of the Carnegie Institution for Science, Pasadena, CA 91101, USA\\
}
\altaffiltext{5}{Author to whom any correspondence should be addressed}

\begin{abstract}

As an evolutionary phase of galaxies, active galactic nuclei (AGN) over a large range of redshift have been utilized for understanding cosmic 
evolution. In particular, the population and evolution of AGNs have been investigated through the study of the cosmic X-ray background in various fields.  
As one of the deep fields observed by \textit{Chandra} with a total of 2.8 Msec exposures, Abell 133 is a special region to investigate AGNs, providing 
a testbed for probing the environmental effect on AGN trigger, since cluster environments can be different from field environments.  
The achieved flux limit of data at 50\%\ completeness level of
$6.95 \times 10^{-16}, 1.43 \times 10^{-16}, \rm \ and \ 1.57 \times 10^{-15}$ \ergs\ $\rm cm^{-2}$ in 0.5-8, 0.5-2, and 2-8 keV.
Using the {\tt wavdetect} and no-source binomial probability (i.e., p<0.007), we analyze the combined \textit{Chandra} image, detecting 
1617 (in 0.5-8 keV),  1324 (in 0.5-2 keV), and 1028 (in 2-8 keV), X-ray point sources in the Abell 133 region. Here, we present the X-ray point source catalogue with the source fluxes, which can be combined with multiwavelength data for future works. We find that the number count distribution of the X-ray point sources is well reproduced with 
a broken power-law model while the best-fit model parameters are sensitive to the fitting range of the number count distribution.
Finally, we find an excess of the number density (decrease of AGN fraction) at the central region of the cluster, which reflects the effect of dense environments on AGN trigger,
as similarly found in studies of other galaxy clusters.

\end{abstract} 
 

\section{INTRODUCTION} \label{section:intro}
 When supermassive black holes (SMBHs) accrete matter and shine as active galactic nuclei (AGN), they release a large amount of energy, which has the potential to influence galaxy evolution (see, e.g., \citealt{Fabian2012} and \citealt{Kormendy2013} for  reviews on AGN feedback and SMBH-galaxy coevolution, respectively). The interplay between SMBHs and galaxies, including triggering AGN activity and feedback to host galaxies may very well  proceed differently depending on the large-scale environments. For example, in dense cluster environments, the gas supply to AGN can be suppressed through environmental effects such as the ram pressure stripping and galaxy harassment (among other mechanisms; e.g., \citealt{Treu2003, Moran2005, Boselli2006}). Such environmental effects can limit the amount of material that is channeled toward the nucleus and accreted onto the SMBH.   
 
Constraints on AGN populations and their evolution can be obtained through the measures of the cumulative number of X-ray sources in the resolved cosmic X-ray background (i.e., $\log N - \log S$; see, e.g., \citealt{Gilli2007,Luo2017}).  Relevant to dense environments, \citet{Ehlert2013}  examined the number of  X-ray point sources  across 43 clusters of galaxies at low redshift ($0.3<z<0.7$) observed with the \textit{Chandra} X-ray telescope. They find that galaxy clusters tend to have a slightly higher number of X-ray point sources than non-cluster environments. Accounting for the much higher galaxy number density in galaxy clusters, this result indicates the lower AGN fraction (i.e., the number of X-ray AGNs divided by the number of galaxies) in the clusters. At the same time, \citet{Ehlert2013} reported that the average AGN fraction in the clusters tends to decrease toward the cluster center, which is consistent with expectations from higher levels of gas removal in denser environments. The environmental effects on AGN trigger has been investigated by various studies using X-ray sources in clusters, groups and fields \citep[e.g.,][]{Arnold2009, Haggard2010, Oh2014}, showing that in general AGN fraction is higher in galaxy groups and fields than in clusters, while the difference of AGN fraction between fields and clusters becomes smaller at higher z \citep[i.e., z$>$1, see][]{Martini2013}.

The nearby galaxy cluster, Abell 133 at $z=0.0566$ has been observed  with the \textit{Chandra} X-ray telescope with a total 2.8 Msec exposure. These observations provide an opportunity to examine the $\log N - \log S$ distribution within a single dense cluster environment over a relatively wide field of view (0.76 deg$^2$) at low redshift. In this paper, we present the detailed analysis of the X-ray point-source population in  the Abell 133 region, and provide a catalog of X-ray point sources suitable for multi wavelength follow-up studies.
In section 2, we describe the galaxy cluster Abell 133 and the archival data. The analysis is described in section 3, and we present our results and discussion in Section 4. We adopt a cosmology of $H_{\rm 0}= 70$ km  s$^{-1}$ Mpc$^{-1}$,  
$\Omega_{\Lambda}=0.7$ and $\Omega_{\rm m}=0.3$.  \\

\section{Archival Chandra Observations} \label{section:data}

We searched the Chandra archive for observations covering Abell 133, and found a total of 2.8 Msec exposures split over 33 separate observations (see Table 1).   All observations were taken with the Advanced CCD Imaging Spectrometer \citep[ACIS;][]{Garmire2003}, with the pointing center placed at the ACIS-I aimpoint (except for ObsID 2203 which was centered on ACIS-S). To cover the large extent of the Abell 133 cluster, which has $R_{200} = 26.6\arcmin$ \citep{Morandi2014}, the majority of \textit{Chandra} observations were taken as a  mosaic  (for reference, the ACIS-I field of view is $16.9\arcmin \times 16.9\arcmin$).  The exposure time of the individual 33 observations ranges from several tens to a few hundreds of ksec as summarized in Table~1. 

We reduced the archival data using the \textit{Chandra} Interactive Analysis of Observations ({\tt CIAO}; \citealt{Fruscione2006}) software version 4.6 with \textit{Chandra} calibration database version 4.6.3. We removed cosmic rays and bad pixels by reprocessing the data with the {\tt Chandra\_repro} script.   After that, we filtered the data for background flares (over 0.5-10 keV) by creating a background image, for which we excluded point sources detected with {\tt celldetect}, and we also excluded the central 3$\arcmin$ (which includes diffuse X-ray emission from hot gas; see below and \S4.2).  We then ran the {\tt deflare} script on the background image (adopting 10 sec binning and a 3$\sigma$ threshold), from which we excluded $\sim1$\% of the total exposure time as intervals with  elevated background count rates.  We next restricted our analysis to a hard 2-8 keV energy band to avoid contamination from hot gas.  We also restricted our analysis to the ACIS S2-4 and ACIS I0-I3 chips; other chips would have  90\% encircled energy fractions (EEFs) $>20 \arcsec$ at 4.510 keV, which is an insufficient spatial resolution. 
Finally, we merged all observations using the {\tt merge\_obs} script, from which we produced the final combined image as shown in Figure~1 which covers $\sim0.8\degr \times 0.85 \degr$ on the sky. In Figure~1, we mark with black dashed circles the $R_{500}$ (inner, 1044 kpc or 15.4\arcmin) and $R_{200}$ (outer, 1596 kpc or 26.6\arcmin) radii, as adopted by \cite{Morandi2014}. These are the radii where the density within the radii are 500 and 200 times of the critical density at the redshift of the cluster, respectively.

Following the method of \cite{Ehlert2013}, we made effective exposure time maps by dividing the exposure area maps of each exposure by its maximum effective area and multiplying by the total exposure time. The combined effective exposure map to the merged 2-8 keV band image is presented in Figure~2. Since the total exposure time is not uniform across the final mosaic image, we present the final survey area as a function of effective exposure time in 2-8 keV band in Figure~3.     
The maximum exposure time is  511.7 ksec, and the 0.55 deg$^2$ out of a total surveys of 0.76 deg$^2$ contains a total exposure larger than 150 ksec.
Even though the mean exposure time is much smaller than that of pencil-beam deep field surveys (e.g., the \textit{Chandra} deeps fields,  CDF-N and CDF-S; \citealt{Alexander2003,Luo2017}), it is comparable to other extended sky surveys such as the \textit{Chandra} multi wavelength project \citep{Kim2007}, and the COSMOS survey \citep{Elvis2009}. 


\begin{figure*}{}
\includegraphics[width = 0.96\textwidth]{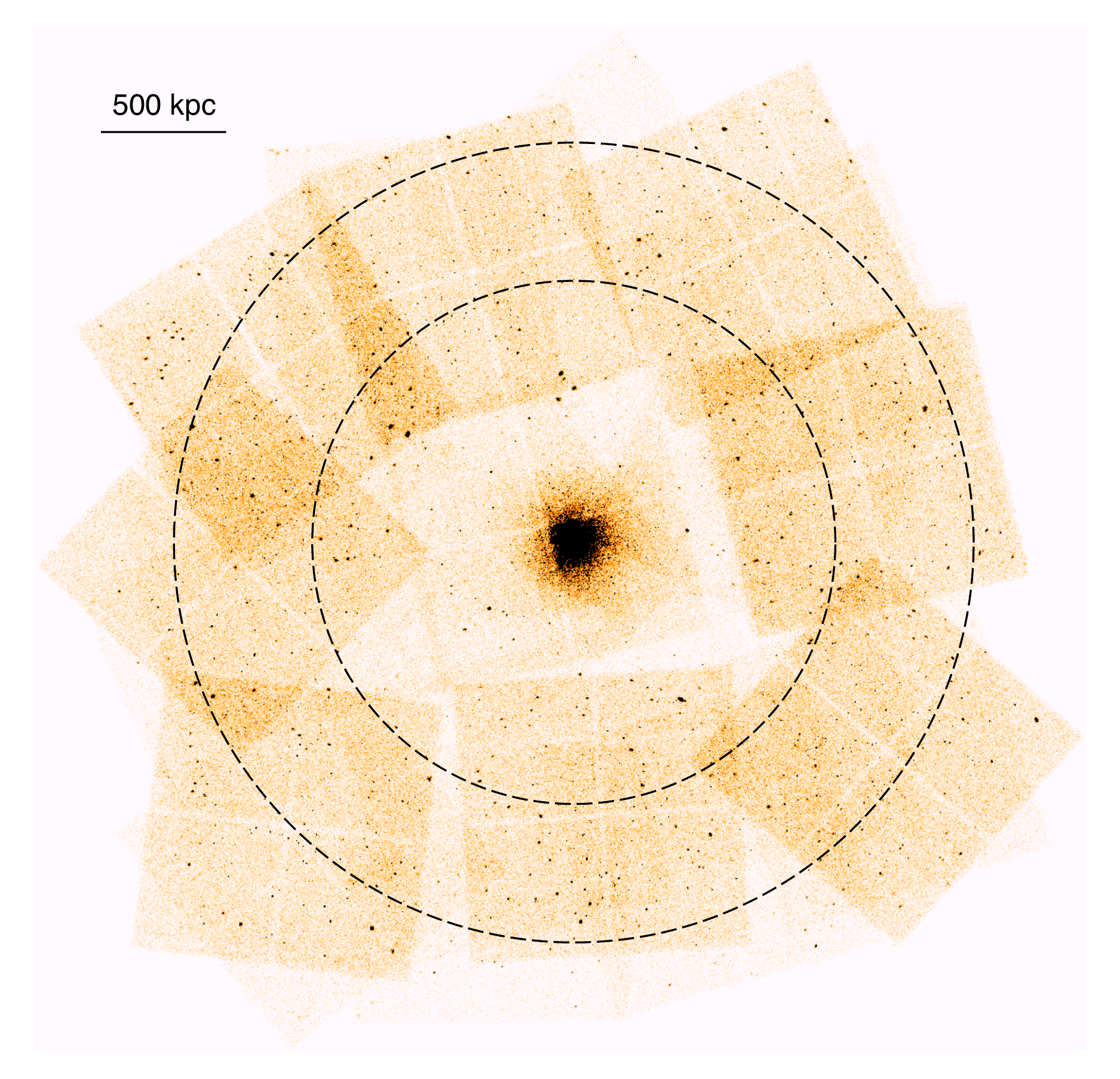}
\caption{
Merged 2-8 keV band image of the Abell 133 region. The image is smoothed with a Gaussian kernel of five pixel radius. $R_{500}$ (1044 kpc or 15.4\arcmin ) and $R_{200}$ (1596 kpc or 26.6\arcmin) are denoted as the inner and outer black dashed lines, respectively.\\
\label{fig:allspec1}}
\end{figure*}

\begin{figure}{}
\centering
\includegraphics[width = 0.44\textwidth]{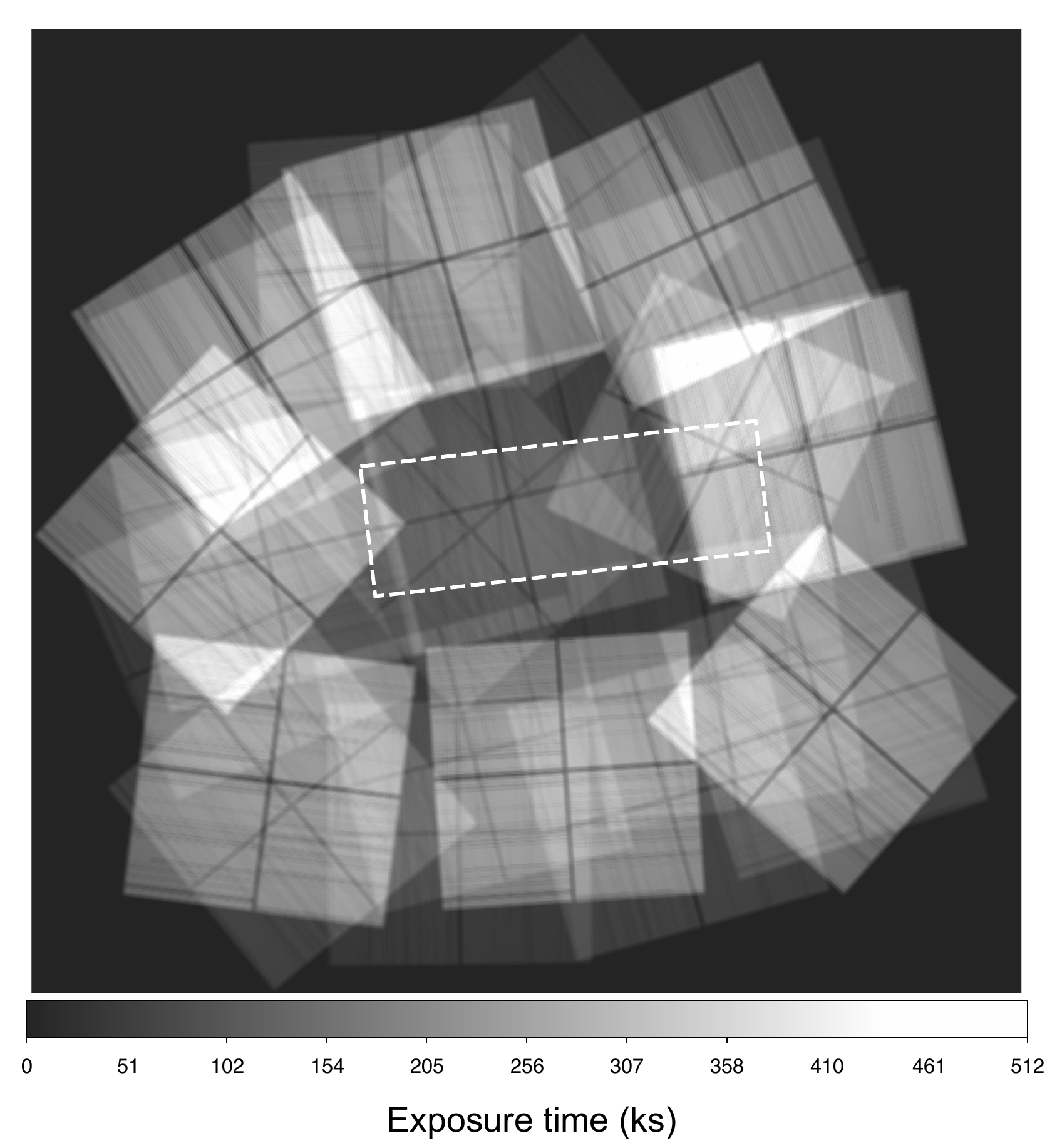}

\caption{
Exposure map with respect to the merged 2-8 keV band image. The white dotted rectangle indicates ACIS-S data (Obs ID: 2203)\\
\label{fig:allspec1}}
\end{figure} 

\begin{figure}{}
\includegraphics[width = 0.44\textwidth]{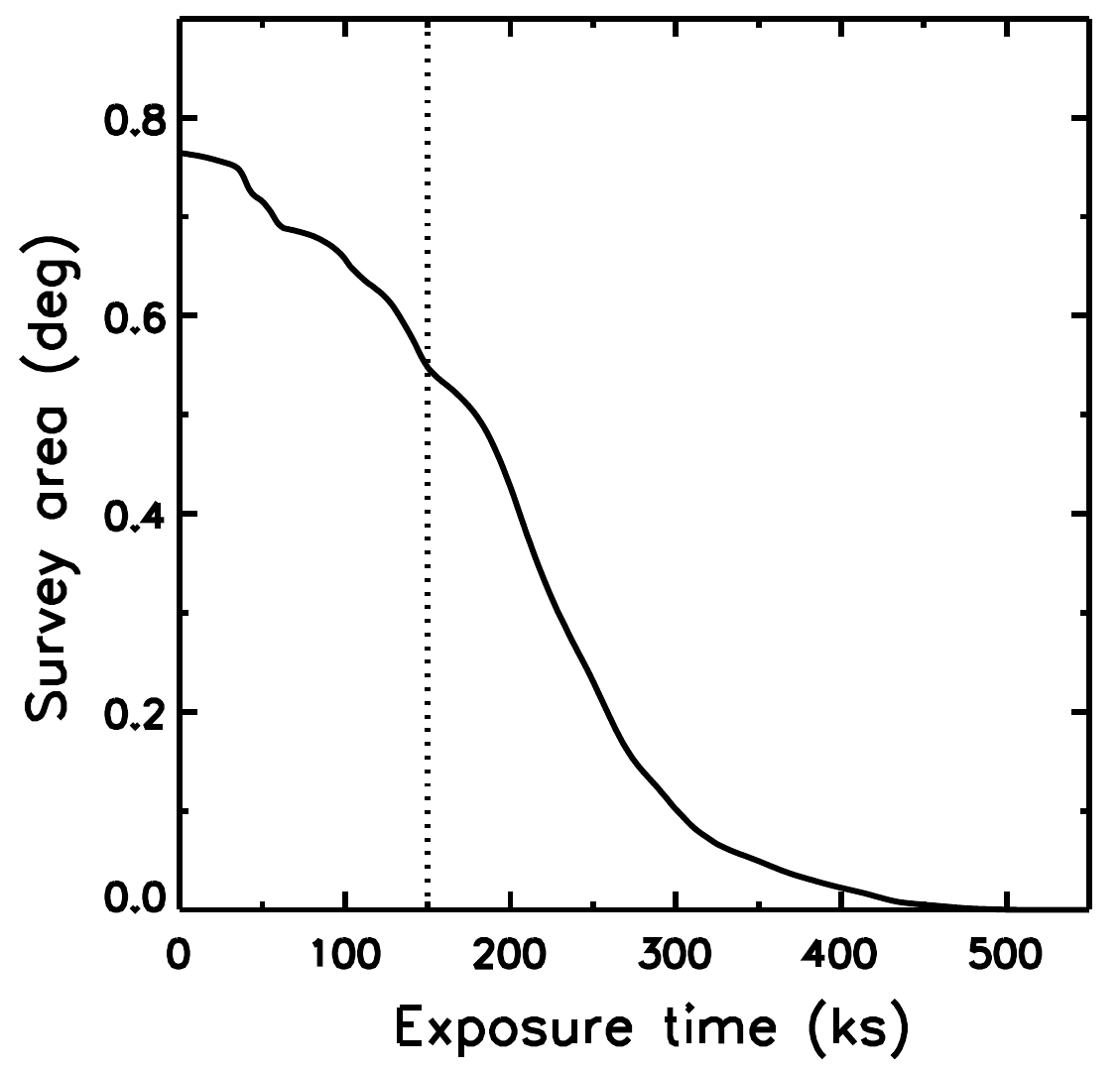}
\caption{
Survey area as a function of effective exposure time in 2-8 keV band.  A total of 72\% of the total survey area (0.55/0.76 deg$^2$) has exposures $>$150 ks (see vertical dotted line). \\
\label{fig:allspec1}}
\end{figure} 

\begin{figure}{}
\includegraphics[width = 0.44\textwidth]{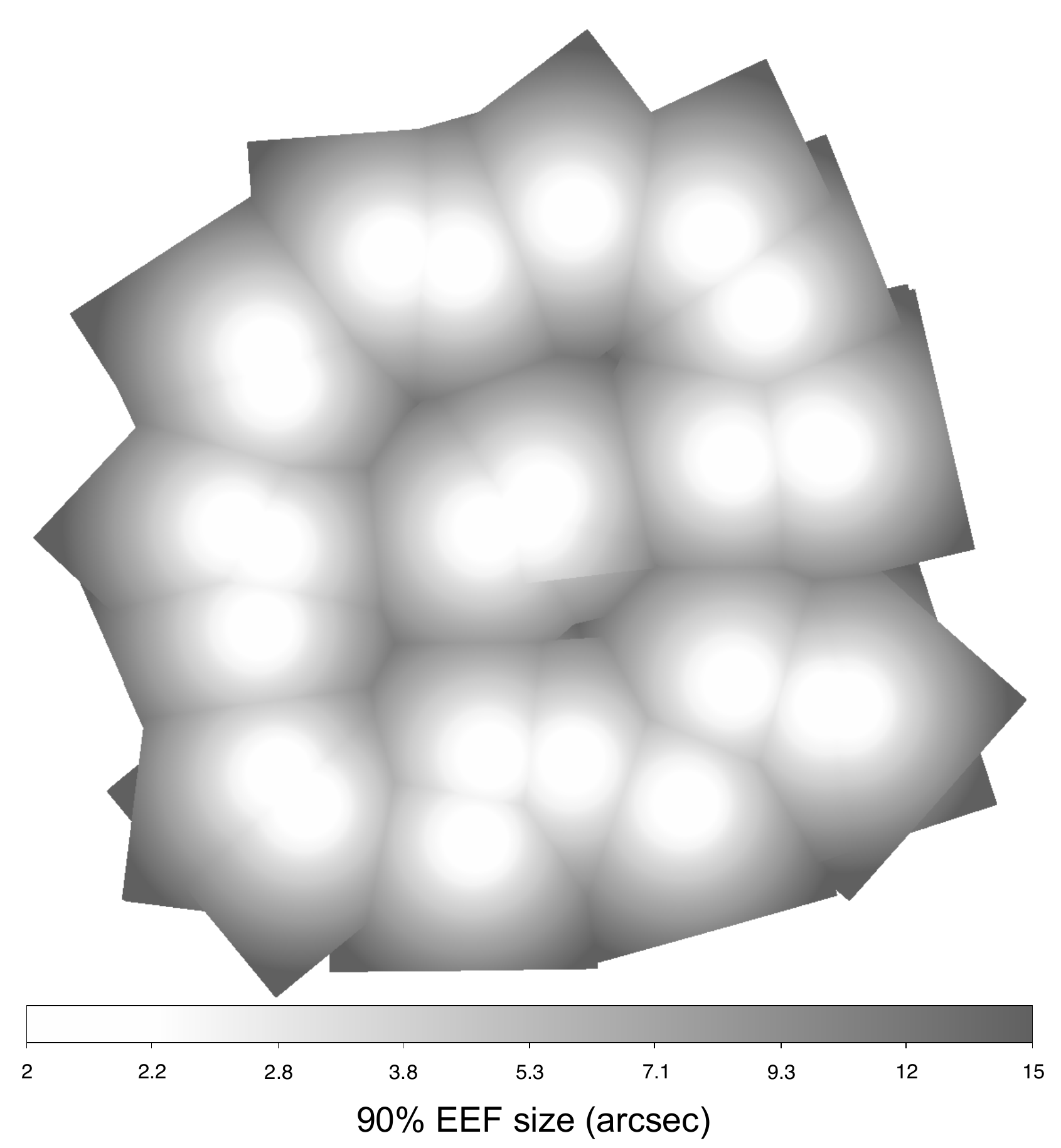}
\caption{
Enclosed energy fraction map at 4.510 keV with respect to the merged 2-8 keV image. 
\label{fig:allspec1}}
\end{figure} 

\begin{deluxetable}{rlrcccc}
\tablewidth{0pt}
\tablecolumns{7}
\tabletypesize{\scriptsize}
\tablecaption{Chandra Observing Log for Abell 133}

\tablehead{
\colhead{Obs ID} &
\colhead{Obs. Start} &
\colhead{Exp} &
\colhead{RA} &
\colhead{DEC} &
\colhead{Obs Cycle} & 
\colhead{Ref.}
\\
\colhead{(1)} &
\colhead{(2)} &
\colhead{(3)} &
\colhead{(4)} &
\colhead{(5)} &
\colhead{(6)} &
\colhead{(7)} 
\\
\colhead{} &
\colhead{} &
\colhead{(ks)} &
\colhead{(deg)} &
\colhead{(deg)} &
}

\startdata
2203		&	2000	Oct	13	&$	35.46	$&$	15.672	$&$	-21.878	$&$	2	$  &  1 \\
3183		&	2002	Jun	24	&$	44.52	$&$	15.670	$&$	-21.878	$&$	3	$  &  2 \\
3710		&	2002	Jun	26	&$	44.59	$&$	15.670	$&$	-21.878	$&$	3	$  &  2 \\
9897		&	2008	Aug	29	&$	69.22	$&$	15.674	$&$	-21.881	$&$	9	$   &  3 \\
12177		&	2010	Aug	31	&$	50.11	$&$	15.453	$&$	-22.062	$&$	11	$ & 4 \\
12178		&	2010	Sep	7	&$	46.83	$&$	15.999	$&$	-22.002	$&$	11	$ & 4 \\\
12179		&	2010	Sep	3	&$	51.10	$&$	15.735	$&$	-22.141	$&$	11	$ & 4 \\\
13391		&	2011	Aug	16	&$	46.43	$&$	15.757	$&$	-22.233	$&$	12	$ & 4 \\\
13392		&	2011	Sep	16	&$	49.89	$&$	15.639	$&$	-21.563	$&$	12	$ & 4 \\\
13442		&	2011	Aug	23	&$	176.69	$&$	15.350	$&$	-21.818	$&$	12	$ & 4 \\\
13443		&	2011	Aug	26	&$	69.68	$&$	15.359	$&$	-21.812	$&$	12	$ & 4 \\\
13444		&	2011	Sep	3	&$	38.26	$&$	15.774	$&$	-21.613	$&$	12	$ & 4 \\\
13445		&	2011	Sep	2	&$	65.18	$&$	15.514	$&$	-22.193	$&$	12	$ & 4 \\\
14333		&	2011	Aug	31	&$	134.76	$&$	15.774	$&$	-21.613	$&$	12	$ & 4 \\\
13446		&	2011	Sep	9	&$	58.42	$&$	15.480	$&$	-21.587	$&$	12	$ & 4 \\\
13447		&	2011	Sep	8	&$	69.13	$&$	15.983	$&$	-21.744	$&$	12	$ & 4 \\\
14338		&	2011	Sep	10	&$	117.50	$&$	15.480	$&$	-21.587	$&$	12	$ & 4 \\\
13448		&	2011	Sep	13	&$	146.12	$&$	15.993	$&$	-21.710	$&$	12	$ & 4 \\\
13449		&	2011	Sep	6	&$	68.15	$&$	15.424	$&$	-21.662	$&$	12	$ & 4 \\\
14343		&	2011	Sep	12	&$	35.30	$&$	15.993	$&$	-21.710	$&$	12	$ & 4 \\\
13451		&	2011	Sep	16	&$	70.12	$&$	15.947	$&$	-22.196	$&$	12	$ & 4 \\\
13452		&	2011	Sep	24	&$	142.07	$&$	15.322	$&$	-22.089	$&$	12	$ & 4 \\\
14345		&	2011	Sep	23	&$	33.74	$&$	15.322	$&$	-22.089	$&$	12	$ & 4 \\\
13454		&	2011	Sep	19	&$	91.79	$&$	16.030	$&$	-21.897	$&$	12	$ & 4 \\\
14346		&	2011	Sep	21	&$	85.90	$&$	16.030	$&$	-21.897	$&$	12	$ & 4 \\\
13456		&	2011	Oct	15	&$	135.63	$&$	15.635	$&$	-22.148	$&$	12	$ & 4 \\\
14354		&	2011	Oct	10	&$	38.64	$&$	15.635	$&$	-22.148	$&$	12	$ & 4 \\\
13518		&	2011	Sep	17	&$	49.60	$&$	15.739	$&$	-21.901	$&$	12	$ & 4 \\\
13450		&	2011	Oct	5	&$	108.18	$&$	15.975	$&$	-22.161	$&$	12	$ & 4 \\\
14347		&	2011	Oct	9	&$	68.69	$&$	15.975	$&$	-22.161	$&$	12	$ & 4 \\\
13453		&	2011	Oct	13	&$	68.95	$&$	15.841	$&$	-21.604	$&$	12	$ & 4 \\\
13455		&	2011	Oct	19	&$	69.63	$&$	15.983	$&$	-21.916	$&$	12	$ & 4 \\\
13457		&	2011	Oct	21	&$	69.13	$&$	15.341	$&$	-22.086	$&$	12	$ & 4	 
\enddata
\label{table:prop}

\tablecomments{
    The `RA' and `Dec' columns indicate the pointing centers of each Chandra observation.}
\tablerefs{
Reference where data were first published: (1) \citet{Fujita2002}; (2) \citet{Vikhlinin2005}; (3) \citet{Sun2009}; (4) \citet{Morandi2014}.  
}
\end{deluxetable}




\section{Analysis} \label{section:analysis}
\subsection{Point source detection}\label{detection}
To detect X-ray point sources in our mosaic image, we first constructed an `EEF map', where each pixel represents the 90\% EEF at 1.5 keV (for soft band) and 4.5 keV (for hard band, see Figure~4). 
The EEF map for the full band is made by averaging our soft- and hard-band maps.
The final EEF was taken as the minimum EEF value of all exposures that contributed to each pixel in the  mosaic, under the assumption that the exposure with the smallest EEF contributed the most weight. Next, we searched for point sources using {\tt wavdetect} \citep{Freeman2002}, adopting wavelet scales of  1, 1.4, 2, 2.8, 4, 5.7, 8, 11.3, and 16, and a threshold significance of $10^{-5}$ (we adopted a relatively high  threshold to avoid rejecting real sources at this point in the analysis, at the cost of increased contamination from 
spurious sources; see, e.g., \citealt{Alexander2001}).  In this process we initially detected 1639 (full band), 1069 (full band), and 1461 (soft band) X-ray point sources.   

To remove false positives from this initial list of X-ray sources, we followed \citet{Lehmer2012} and \cite{Ehlert2013} and calculated the probability that a detected source was spurious via the no-source binomial probability:
\begin{equation}
P = \sum\limits_{x \ge S}^{N} \frac{N!}{x! (N-x)!} p^x (1-p)^{n-x},
\end{equation}
 where $N$ is the total number of source and background counts within a source aperture, $S$ is the number of source counts in that aperture, and $p=1/(1+\rm \Omega_{bkg}/\Omega_{src}$), where $\rm \Omega_{src}$ and $\rm \Omega_{bkg}$ are the areas of  source and background apertures, respectively. For example, we identified 38, 50, and 87 sources in the hard band as spurious when we used $P>0.007, 0.004,$ and $0.001$, respectively. In the previous work by \citet[e.g.,][]{Luo2017}, the value of the no-source binomial probability was empirically determined to balance high reliability (minimizing the number of spurious targets) and high completeness (large number of targets)  based on multiwavelength information. Since we do not have multiwavelength constraints on the potential X-ray sources, we adopted $P>0.007$ to remove false positives by following \citet{Luo2017} and the final catalog contains 1617 (full band), 1028(hard band), and 1324 (soft band) X-ray point sources.

\subsection{Astrometry correction}\label{astrometry correction}
To improve the accuracy of the positions of the detected X-ray point sources, we applied an astrometry correction by searching for optical counterparts, using archival images from the Canada-France-Hawaii telescope (CFHT). Eleven g-band images were taken as a mosaic to cover 1 square degree with the total exposure time of 1320s and 5$\sigma$ limiting magnitude is 26.5. We cross-matched the 1028 X-ray point sources detected in hard band with the list of 694 optical point sources with $g$-band magnitudes brighter than 19 mag, using the {\tt CIAO} tool {\tt wcs\_match} with a 3\arcsec\ matching radius and 0$\farcs$6  residual limit.  We identified 30 optical counterparts, with which we used the tool {\tt wcs\_update} to update the X-ray astrometry, which included shifts in the x (east-west) and y (south-north) directions by -1.1  and -0.1  pixels, respectively, and in rotation by 0.01$\arcdeg$ (counter-clockwise). We note that we tried to correct the astrometry for individual exposures (prior to merging) similarly to several prior works for deep fields \citep[e.g.,][]{Luo2017}. However, we found that the astrometry correction for individual exposure introduced additional uncertainties for our dataset mainly due to the lower exposure time of each exposure. Thus, we decided to correct the astrometry after merging. We do not discuss optical properties in this paper.  However, for reference,  we  tabulate available information on matched optical counterparts in our final X-ray catalog  (see Section 3.4).\\

\subsection{Flux estimation}\label{astrometry correction}
To measure the X-ray flux of each point source, we adopted a source aperture as the radius set to the value of the 90\% EEF at the position of each source.  The number of background counts per pixel was estimated using a background image for which we excised all detected sources identified by {\tt wavdetect} (i.e., including spurious sources), using the annuli with inner and outer radii between 2 and 3 times the 90\% EEF.  We then calculated the net number of counts in each source aperture, to which we applied a 90\% aperture correction. Corresponding error is estimated at the 90\% confidence level, assuming Poisson statistics \citep{Gehrels1986}.  

 We converted the net count rates to fluxes using  conversion factors calculated from {\tt WEBPIMMS}\footnote{http://cxc.harvard.edu/toolkit/pimms.jsp}. For the conversion factor, we adopted the \textit{Chandra} Cycle 12 effective area curves (the vast majority of our observations were taken in cycle 12), and a power-law model with photon index $\Gamma=1.4$ (i.e., $N_E \propto E^{-1.4}$, where $N_E$ is the photon flux density and $E$ is the photon energy) and foreground Galactic absorption with a column density $N_{\rm H} = 1.6 \times 10^{20}\, {\rm cm^{-2}}$.   Our adopted column density was determined as the average value from the \citet{Dickey1990} \ion{H}{1} maps, which range from 1.4-1.6$\times 10^{20} \rm\ cm^{-2}$ across the field of view of our \textit{Chandra} mosaic.   We note that the effects from our choices for Cycle number and column density are not significant, only introducing $\lesssim$ 5\% and 1\% systematic errors from Cycle number and column density, respectively.  To estimate unabsorbed fluxes at each band, we obtained a conversion factor of $1.233 \times 10^{-11}$ , $6.109 \times 10^{-12}$ and  $2.214\times 10^{-11}\, \rm erg/s^{}$/cm$^{2}$ per 1 count$^{}$/s$^{}$,  for the full-, soft- and hard- bands respectively.
 
Since we have no information on the distance to individual X-ray sources, we are not able to calculate X-ray luminosities for each source, nor can we ascertain on an individual basis if the X-ray sources are associated with the cluster (or if they are  foreground/background sources). In the hard band, the minimum flux among the detected targets is $4.3 \times 10^{-16}\, \rm erg\ s^{-1} cm^{-2}$, which would correspond to a luminosity of 7.9 $\times 10^{38}\, \rm erg\ s^{-1}$ at the distance of Abell 133. At such low luminosities, there could be contaminations from low mass X-ray binaries and high mass X-ray binaries within the 90\% \textit{Chandra} EEF  \citep[if at the cluster redshift; e.g.,][]{Grimm2003,Gilfanov2004,Mineo2012,Lehmer2014}.  However, due to the lack of the distance information,  we make no attempt to estimate X-ray binary contribution fractions, as such an investigation would require further multi-wavelength data (e.g., optical spectra to determine redshifts of optical counterparts). \\

\begin{table*}
\begin{longtable*}{cccccccccccc}
\caption{Main {\it Chandra} source catalog}\\
\hline
& \multicolumn{2}{c}{X-ray coordinates}  & 
&&
&  & \multicolumn{2}{c}{Optical coordinates}
&&  &\\

\multicolumn{1}{c}{\#} & \multicolumn{1}{c}{RA} & \multicolumn{1}{c}{DEC} 
&\multicolumn{1}{c}{FB} &\multicolumn{1}{c}{SB} &\multicolumn{1}{c}{HB}
& \multicolumn{1}{c}{HR} & \multicolumn{1}{c}{RA} & \multicolumn{1}{c}{Dec}
& \multicolumn{1}{c}{$g$ mag} & \multicolumn{1}{c}{$g-r$} & \multicolumn{1}{c}{$\alpha_{ox}$} \\

%
\multicolumn{1}{c}{(1)} & \multicolumn{1}{c}{(2)}  & \multicolumn{1}{c}{(3)} &
\multicolumn{1}{c}{(4)} & \multicolumn{1}{c}{(5)} & \multicolumn{1}{c}{(6)} &
\multicolumn{1}{c}{(7)} & \multicolumn{1}{c}{(8)} & \multicolumn{1}{c}{(9)} &
\multicolumn{1}{c}{(10)} & \multicolumn{1}{c}{(11)} & \multicolumn{1}{c}{(12)}  \\

\hline
\endfirsthead
\endhead
\endfoot
\hline
\multicolumn{12}{l}{\scriptsize{ Note -- Col. (2-3): Right ascension and Declination of X-ray point source. Col. (4-6): Net counts in full- (0.5-8 keV), soft- (0.5-2 keV), and hard-(2-8 keV)}} \\
\multicolumn{12}{l}{\scriptsize{band. The conversion factors from the net counts to the fluxes are $1.233 \times 10^{-11}$, $6.109 \times 10^{-12}$, and $2.214\times 10^{-11}\, \rm erg/s^{}$/cm$^{2}$ per 1 count$^{}$/s$^{}$ for the full-, soft- }}\\
\multicolumn{12}{l}{\scriptsize{and hard- bands respectively. Col. (7): Hardness ratio. Col. (8-9): Right ascension and Declination of optical counterpart. Col. (10): $g$-magnitude. Col. (11): }}\\
\multicolumn{12}{l}{\scriptsize{$g-r$ color. Col. (12): optical to X-ray power-law slope. Table 2 is published in its entirety in the machine-readable format. A portion is shown here for guidance}} \\
\multicolumn{12}{l}{\scriptsize{regarding its form and content.}} \\

\endlastfoot
    1 &  15.17015  &-22.05955  &$	1751.5  _{ -68.3  }^{  +70.4 }$&	$ 1217.0 _{  -56.8  }^{  +59.0 }$&$  490.3 _{  -35.8}^{	    +38.0}$&	$   -0.43  	_{	 -0.04  	}^{  +0.04  }$&		 15.17018 & -22.05977 &  19.30 &   0.04 &   1.43\\  
    2 &  15.18871  &-22.06865  &$	  119.2   _{-17.4   }^{ +19.6  }$&	$  69.8  _{ -13.2   }^{ +15.4  } $&$  25.4  _{  -7.7 }^{	    +10.0 }$&	$  -0.47  	_{	 -0.18  	}^{  +0.21 }$  & --- &--- &---&---&---\\                                             
    3  & 15.19384  &-22.04413  &$	  99.1   _{-15.8  }^{  +18.0 }$&	$   26.8  _{  -7.9   }^{  +10.2   }$&$46.9  _{ -10.7  }^{	  +12.9}$&	$    0.27 	_{	  -0.19 	}^{   +0.23  }$  & --- &--- &---&---&---\\                                             
    4  & 15.19691  &-21.89512  &$	 349.3  _{ -30.2 }^{   +32.3}$&	$   173.0 _{  -21.0 }^{   +23.2 }$&$  105.0 _{  -16.3 }^{	   +18.5}$&	$   -0.24  	_{	 -0.10   	}^{ +0.11   }$  & --- &--- &---&---&---\\                                             
    5  & 15.20327  &-22.02950  &$	  61.6  _{ -12.3  }^{  +14.6 }$&	$    9.6    _{-4.5   }^{  +6.9  }    $&$  29.7  _{  -8.4 }^{	   +10.6}$&	$    0.51  	_{	 -0.27  	}^{  +0.36   }$                         & --- &--- &---&---&---

\end{longtable*}
\end{table*}

\subsection{Main {\it Chandra} source catalog}\label{astrometry correction}
Using the detected sources in three bands, we construct a main {\it Chandra} source catalog.
We cross-matched the detected point sources in the three bands using matching radii of 2.5\arcsec\ for point sources located within 6\arcmin\ from the closest aimpoint (see e.g., Figure~4) and 4\arcsec\ for the others. The main {\it Chandra} source catalog includes 1952 point sources, which are detected in at least one of the three bands (Table~2). We estimated net counts for each band with the uncertainty of 90\% confidence level, and upper limits are at 95\% confidence (2$\sigma$).  For sources detected in both soft- and hard-bands, we calculated hardness ratios: 
\begin{equation}
{\rm Hardness\ ratio} =  \frac{H-S}{H+S},
\end{equation}
where $H$ and $S$ are the number counts in hard band and soft bands, respectively. In Table~2, RA and Dec information is adopted from the order of hard-, full-, and soft- bands. 

We also provide optical information for any X-ray point source that has an optical counterpart.  To identify optical counterparts, we consider 15093 optical points sources with $r_{\rm mag} < 22.5$ in our CFHT images   Out of 1951 X-ray detections, we found optical counterparts for 310 sources adopting a matching radius of 1.5\arcsec\ which is adopted in e.g., \cite{Xue2011,Luo2017}. In Table~2, we present their optical positions, the $g$ magnitude, $g-r$ color, and optical to X-ray spectral index $\alpha_{\rm ox}$=-0.384 ${\rm log}\ (f_{2\rm kev}/f_{2500{\rm \angstrom}})$ \citep{Tananbaum1979}. We note that $f_{2\rm kev}$ was converted from full band flux, assuming a photon index $\Gamma=1.4$ and $f_{2500{\rm \angstrom}}$ was estimated by extrapolating from $g$ magnitude, assuming $\alpha_{\rm \lambda}=-1.56$ \citep{Vandenberk2001}.

\begin{figure}{}
\includegraphics[width = 0.44\textwidth]{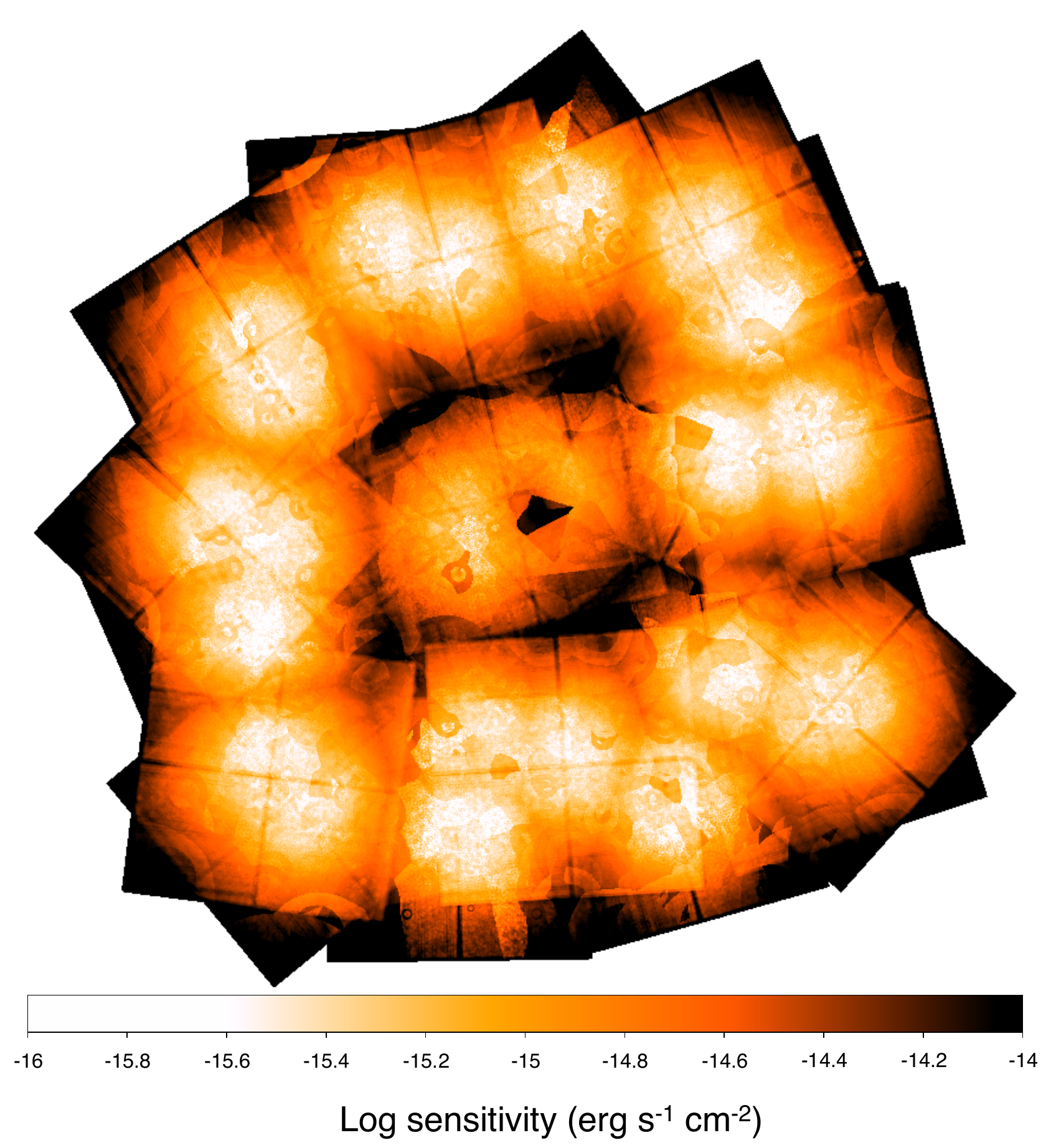}
\caption{
The sensitivity map with respect to the merged 2-8 keV image.
\label{fig:allspec1}}
\end{figure} 

\subsection{Sensitivity map}\label{astrometry correction}
To understand the completeness of our survey we constructed a sensitivity map for each band, following an algorithm outlined by \citet{Ehlert2013}. Our sensitivity map provides the minimum number of photons required for a detection at every pixel position,  with respect to the background. We first made a background image by excising regions containing X-ray point sources (masking twice the 90\% EEF size, using the {\tt CIAO} tool {\tt roi}), and we then randomly replaced pixel values within each excised region assuming Poisson noise, based on the distribution of counts in  annuli with radii spanning 2-3$\times$ the  90\% EEF (using the tool {\tt dmfilth}).    

To build the sensitivity map, we defined local and global backgrounds at the positions of every pixel.   The local backgrounds were calculated from the background image described above, using annuli centered on each pixel with inner and outer radii of 2 and 3 times the 90\% EEF, respectively. For the global backgrounds, we utilized the merged image (i.e., including the X-ray point sources), from which we adopted the maximum number of counts for any source that falls within an annulus spanning 10-20 times the 90\% EEF.  If there were no point source within that annulus, then we adopted as the global background the number of counts from the closest point source.  By using this strategy, we can minimize the p value and  the photon number counts required for a detection. In  Equation 1, we then adopted $S$ as the number of the local background counts, $N$ as the total number of the local and global background counts, and $p=1/(1+\rm \Omega_{global}/\Omega_{local}$), where $\rm \Omega_{local}$ and $\rm \Omega_{global}$ are the areas of the local background and the global background, respectively. We then calculated the minimum number of counts ($x$) to yield $P<0.007$, which represents the minimum number of counts required for a detection at each pixel.  From these  count limits,  the exposure map, and the previously determined conversion factors, we derived the flux sensitivity map and the map in the hard band is shown in Figure~5.  Note that the sensitivity varies dramatically from $\sim 10^{-16}\ {\rm to}\ 10^{-13}\ \rm erg\ s^{-1}\ cm^{-2}$ across the survey field of view.

Figure~6 shows the cumulative survey area as a function of sensitivity, and we tabulate various completeness flux thresholds from 90\% completeness down to 20\% in Table~3.  We also quote corresponding luminosities at the distance of Abell 133 in Table~3, and the number of detected X-ray point sources above each flux threshold. Through the rest of this paper, particularly in \S4.2, we define high- and low-flux subsamples based on a  threshold of the 80\% completeness level. \\

\begin{figure}{}
\includegraphics[width = 0.44\textwidth]{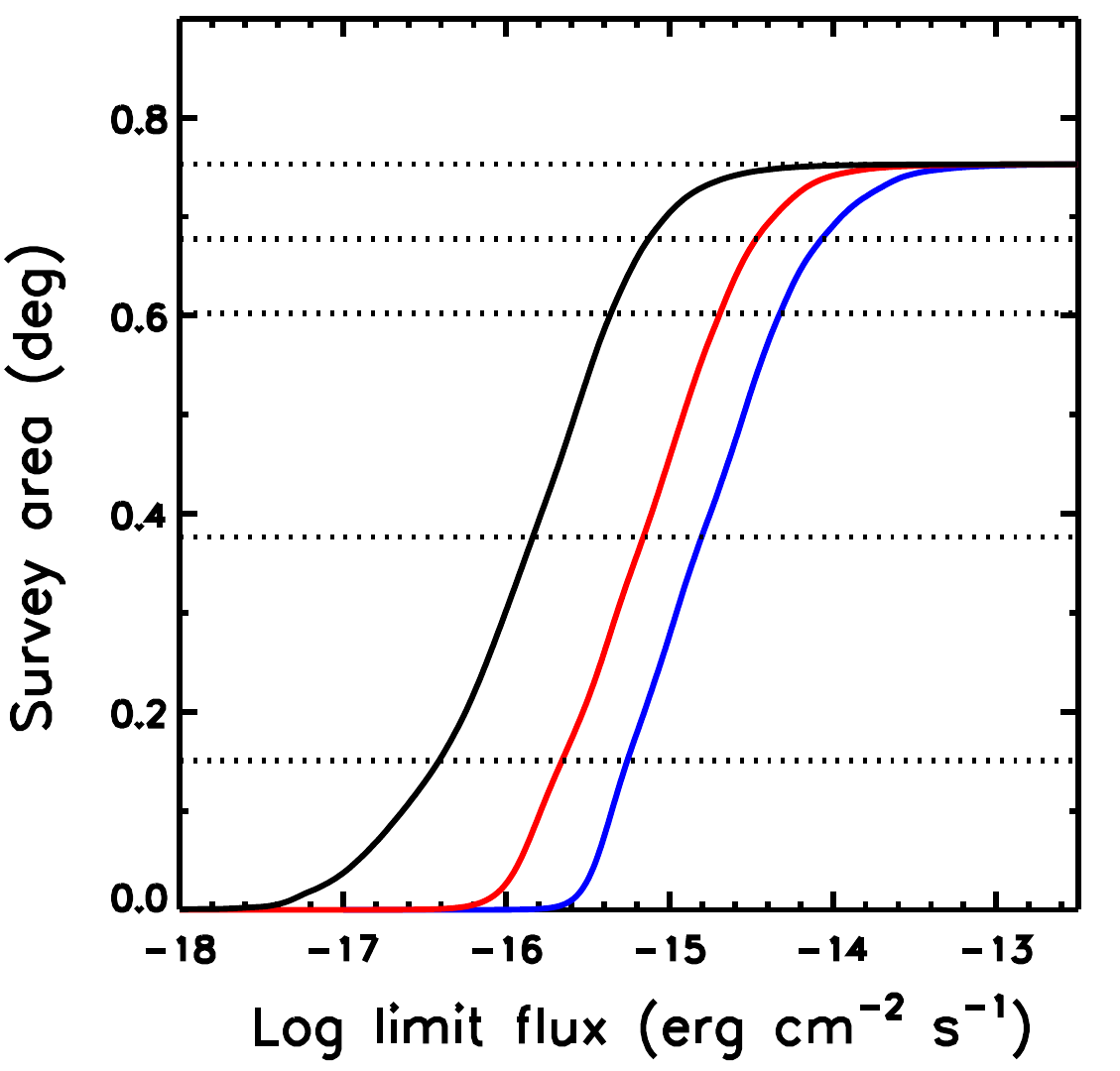}
\caption{
Survey area as a function of the flux detection sensitivity limit. Horizontal dotted lines mark the flux limits for 100\%, 90\%, 80\%, 50\%, and 20\% sky coverage (from top to bottom).  
{\bf Black, red, and blue indicate soft-, full-, and hard- bands}, respectively.
\label{fig:allspec1}}
\end{figure}

\begin{deluxetable}{cccc} 
\tablewidth{0pt}
\tablecolumns{6}
\tabletypesize{\scriptsize}
\tablecaption{Flux Limits and Completeness}

\tablehead{
\colhead{Completeness (\%)} &
\colhead{$\rm f_{0.5-8} keV $} &
\colhead{$\rm f_{0.5-2} keV $} &
\colhead{$\rm f_{2-8} keV $} 
\\
\colhead{(1)} &
\colhead{(2)} &
\colhead{(3)} &
\colhead{(4)} 
\\
\colhead{} &
\colhead{erg $\rm cm^{-2} \ s^{-1}$} &
\colhead{erg $\rm cm^{-2} \ s^{-1}$} &
\colhead{erg $\rm cm^{-2} \ s^{-1}$} }
\startdata
90 &$3.38\times 10^{-15}$& $7.08 \times 10^{-16}$&           $8.50 \times 10^{-15}$  \\
80 &$2.04\times 10^{-15}$ &$4.26 \times 10^{-16}$&     $4.75 \times 10^{-15}$  \\
50 &$6.95\times 10^{-16}$&$1.43 \times 10^{-16}$&       $1.57 \times 10^{-15}$   \\
20 &$2.20\times 10^{-16}$ & $3.83 \times 10^{-17}$ &       $5.55 \times 10^{-16}$    
\enddata
\label{table:prop}


\tablecomments{
    Col. (1): completeness level. Col. (2-4): flux limit for each band for given completeness.}
\end{deluxetable}

\begin{figure}{}
\centering
\includegraphics[width = 0.44\textwidth]{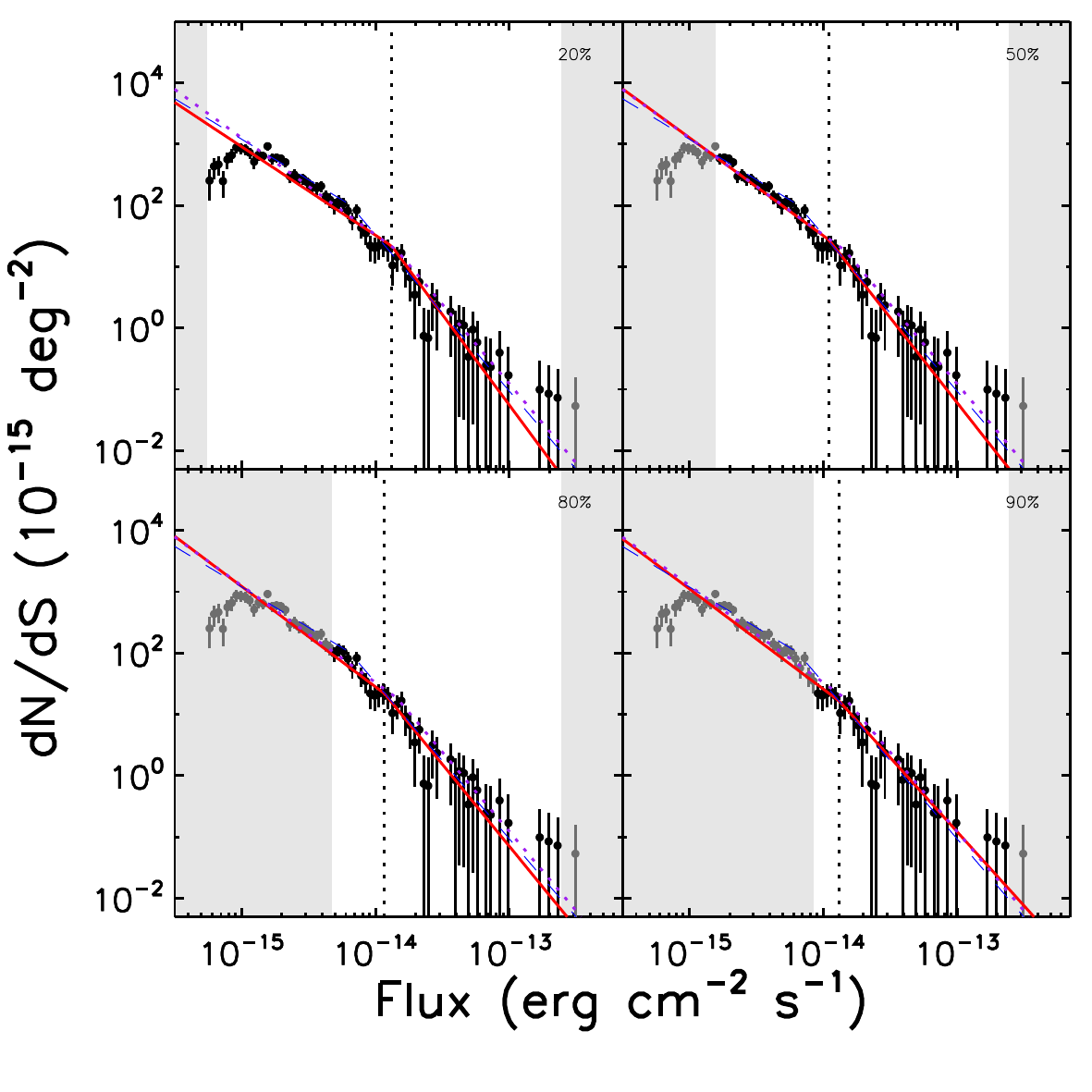}
\caption{
Differential number count distribution in the hard band with various fitting ranges. Black circles denote data points included in the model fits, and grey circles in each grey shaded region are omitted.  The red solid line indicates the best-fit to our broken power laws model. Error bars indicate 1sigma confidence levels.  The blue dashed and purple dotted lines show the results for the ChaMP field \citep{Kim2007} and the CDF-S field \citep{Lehmer2012}, respectively.  The vertical dotted line denotes the break flux. 
\label{fig:allspec1}}
\end{figure} 

\begin{figure}{}
\centering
\includegraphics[width = 0.44\textwidth]{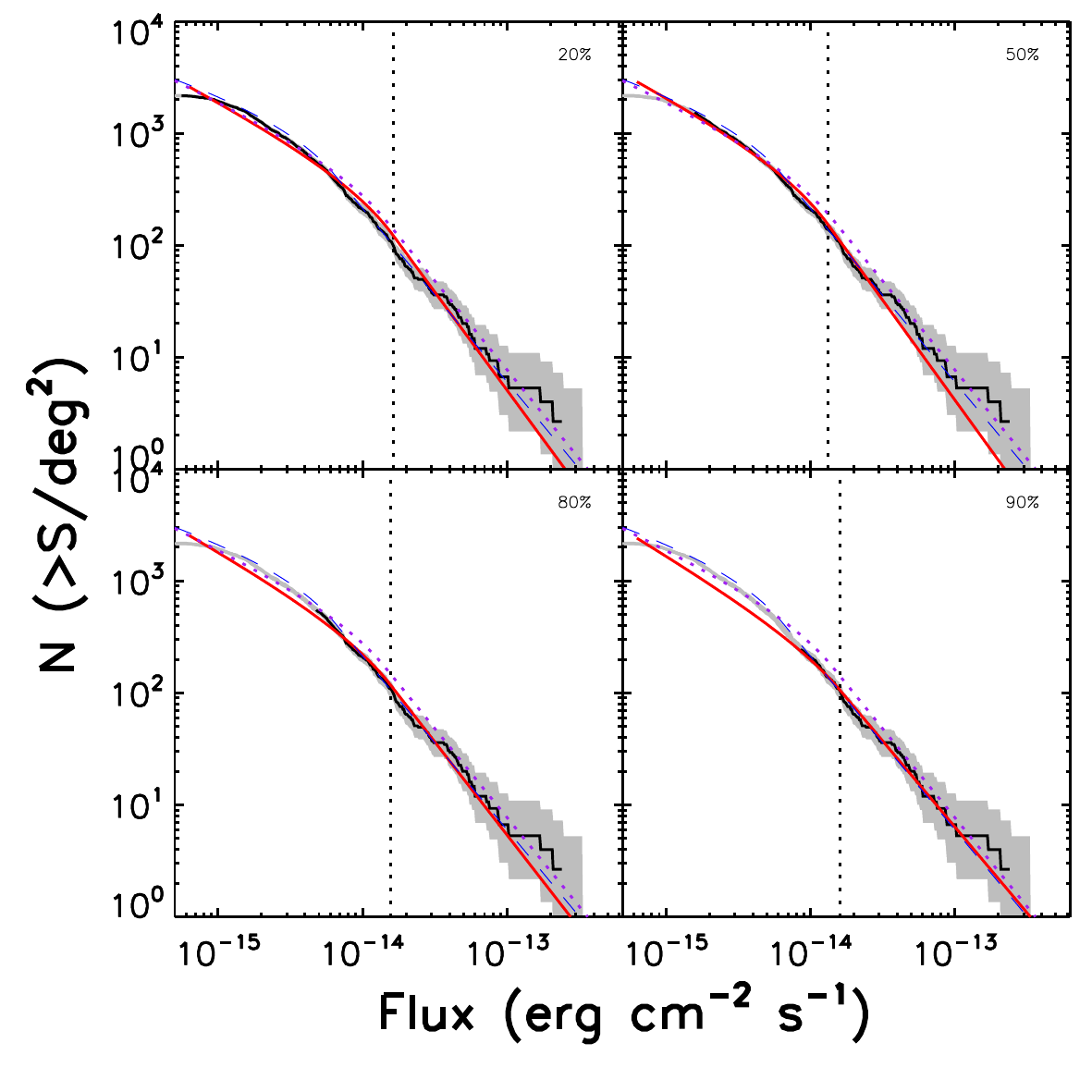}
\caption{
Cumulative number count distribution in the hard band with various fitting ranges. Black- and grey solid shows the
our result within and out of fitting range.
The grey shaded area indicates the 1$\sigma$ confidence level. 
Other symbols have the same meaning as in Figure~7.
\label{fig:allspec1}}
\end{figure}

\begin{deluxetable*}{ccccccc} 
\tablewidth{0pt}
\tablecolumns{6}
\tabletypesize{\scriptsize}
\tablecaption{Model parameters for number distribution}

\tablehead{
\colhead{Completeness (\%)} &
\colhead{$\rm flux limit$} &
\colhead{$K$} &
\colhead{$\gamma_{1}$} &
\colhead{$\gamma_{2}$}  &
\colhead{$S_{b}$}
\\
\colhead{} &
\colhead{(erg $\rm cm^{-2} \ s^{-1}$)} &
\colhead{} &
\colhead{}&
\colhead{}&
\colhead{}
\\
\colhead{(1)} &
\colhead{(2)} &
\colhead{(3)} &
\colhead{(4)}  &
\colhead{(5)} &
\colhead{(6)} } 
\startdata
\noalign{\vskip 1mm} 
\multicolumn{6}{c}{Full band} \\
\noalign{\vskip 1mm} 

\hline
\hline
\multicolumn{6}{c}{Cumulative number count distribution} \\
\noalign{\vskip 1mm} 
20 & $2.20 \times 10^{-16}$ & $    951 \pm      41 $&$        1.32 \pm       0.04 $&$        2.91 \pm       0.14 $&$    18.9 \pm   1.0 $\\
50 & $6.95 \times 10^{-16}$ & $    1470 \pm      19 $&$        1.51 \pm       0.01 $&$        2.40 \pm       0.03 $&$    13.2 \pm   0.3 $\\ 
80 & $2.04 \times 10^{-15}$ & $    1629 \pm      21 $&$        1.54 \pm       0.01 $&$        2.54 \pm       0.02 $&$    14.4 \pm   0.4 $\\ 
90 & $3.38 \times 10^{-15}$   &$    1618 \pm      28 $&$        1.55 \pm       0.01 $&$        2.52 \pm       0.02 $&$    14.6 \pm   0.4 $\\ 
\hline
\noalign{\vskip 1mm}    
\multicolumn{6}{c}{Differential number count distribution} \\
\noalign{\vskip 1mm} 
20 & $2.20 \times 10^{-16}$ & $     577 \pm      53 $&$        0.99 \pm       0.10 $&$        3.02 \pm       0.41 $&$    11.9 \pm   2.4 $\\ 
50 & $6.95 \times 10^{-16}$ & $    753 \pm      57 $&$        1.11 \pm       0.05 $&$        2.75 \pm       0.26 $&$    10.0 \pm   1.8 $\\ 
80 & $2.04 \times 10^{-15}$ & $   688 \pm      68 $&$        1.08 \pm       0.05 $&$        2.79 \pm       0.28 $&$    10.3 \pm   1.9 $\\ 
90 & $3.38 \times 10^{-15}$   &$     689 \pm      62 $&$        1.11 \pm       0.06 $&$        2.70 \pm       0.26 $&$    10.4 \pm   1.9 $\\ 
\hline
\noalign{\vskip 1mm} 
\multicolumn{6}{c}{Soft band} \\
\noalign{\vskip 1mm} 

\hline
\hline
\multicolumn{6}{c}{Cumulative number count distribution} \\
\noalign{\vskip 1mm} 
20 & $3.83 \times 10^{-17}$ & $    584 \pm       7 $&$        0.96 \pm       0.01 $&$        2.24 \pm       0.02 $&$    1.9 \pm   0.0 $\\ 
50 & $1.43 \times 10^{-16}$ & $   585 \pm       5 $&$        1.20 \pm       0.02 $&$        2.12 \pm       0.01 $&$    1.8 \pm   0.0 $\\ 
80 & $4.26 \times 10^{-16}$ & $    632 \pm       10 $&$        1.36 \pm       0.05 $&$        2.07 \pm       0.02 $&$    1.6 \pm   0.0 $\\  
90 & $7.08 \times 10^{-16}$   &$    659 \pm      12 $&$        1.37 \pm       0.04 $&$        2.08 \pm       0.02 $&$    1.6 \pm   0.0 $\\ 
\hline
\noalign{\vskip 1mm}    
\multicolumn{6}{c}{Differential number count distribution} \\
\noalign{\vskip 1mm} 
20 & $3.83 \times 10^{-17}$ & $     729 \pm     128 $&$        0.60 \pm       0.14 $&$        1.89 \pm       0.08 $&$    0.8 \pm   0.1 $\\ 
50 & $1.43 \times 10^{-16}$ & $     607 \pm      44 $&$        1.09 \pm       0.08 $&$        1.87 \pm       0.07 $&$    0.9 \pm   0.1 $\\ 
80 & $4.26 \times 10^{-16}$ & $  637 \pm     103 $&$        1.16 \pm       0.24 $&$        1.88 \pm       0.07 $&$    0.8 \pm   0.1 $\\ 
90 & $7.08 \times 10^{-16}$   &$      664 \pm      89 $&$        1.40 \pm       0.28 $&$        1.89 \pm       0.06 $&$    0.7 \pm   0.1 $\\  
\hline
\noalign{\vskip 1mm} 
\multicolumn{6}{c}{Hard band} \\
\noalign{\vskip 1mm} 

\hline
\hline
\multicolumn{6}{c}{Cumulative number count distribution} \\
\noalign{\vskip 1mm} 
20 & $5.55 \times 10^{-16}$ & $    1370 \pm      20 $&$        1.67 \pm       0.02 $&$        2.75 \pm       0.07 $&$    16.3 \pm   0.6 $\\ 
50 & $1.57 \times 10^{-15}$ & $       1543 \pm      41 $&$        1.67 \pm       0.01 $&$        2.78 \pm       0.05 $&$    13.2 \pm   0.8 $\\ 
80 & $4.75 \times 10^{-15}$ & $  1393 \pm      47 $&$        1.71 \pm       0.01 $&$        2.66 \pm       0.03 $&$    15.7 \pm   0.8 $\\ 
90 & $8.50 \times 10^{-15}$   &$       1354 \pm      50 $&$        1.76 \pm       0.01 $&$        2.54 \pm       0.03 $&$    16.1 \pm   0.7 $\\
\hline
\noalign{\vskip 1mm}    
\multicolumn{6}{c}{Differential number count distribution} \\
\noalign{\vskip 1mm} 
20 & $5.55 \times 10^{-16}$ & $    903 \pm     159 $&$        1.44 \pm       0.12 $&$        2.95 \pm       0.64 $&$    13.1 \pm   2.5 $\\ 
50 & $1.57 \times 10^{-15}$ & $    1265 \pm      79 $&$        1.60 \pm       0.04 $&$        2.77 \pm       0.34 $&$    11.0 \pm   1.9 $\\ 
80 & $4.75 \times 10^{-15}$ & $  1218 \pm     118 $&$        1.64 \pm       0.05 $&$        2.68 \pm       0.36 $&$    11.7 \pm   2.1 $\\
90 & $8.50 \times 10^{-15}$   &$     1112 \pm     713 $&$        1.63 \pm       0.27 $&$        2.42 \pm       0.50 $&$    13.1 \pm   7.9 $
   
\enddata
\label{table:prop}

\tablecomments{
    Col. (1): completeness level. Col. (2): flux limit for given completeness. Col. (3): 
   normalization factor. Col. (4): faint end slope. Col. (5): bright end slope. 
   Col. (6): break flux in unit of $10^{-15} \rm erg\ s^{-1} cm^{-2}$.}
\end{deluxetable*}

 

\section{Results and discussion} \label{section:result}
\subsection{Number count distribution}\label{astrometry correction}
For our detected X-ray point sources, we investigate cumulative and differential number counts.
The cumulative number counts above the given flux, $S$, is

\begin{equation}
N(>S) = \sum\limits_{S_{i} \ge S}  \frac{1}{\Omega_{i}},
\end{equation}
where the $\Omega_{i}$ is the survey area at the given flux of the $i$th source. 

The differential number counts is the derivative form of cumulative
number counts, and can be calculated as follows:

\begin{equation}
\frac{dN}{dS} \bigg \rvert_{i}=-\frac{N_{i+1}-N_{i}}{S_{i+1}-S_{i}},
\end{equation}

The distributions for X-ray point sources in previous works are represented by a double (or broken) 
power law \citep[see e.g.,][]{Kim2007,Lehmer2012,Luo2017}. We adopt the same equations here as \cite{Kim2007}: 

\begin{equation}
\frac{dN}{dS} =\begin{cases}
-K(S/S_{ref})^{-\gamma_{1}},  & \text{$S <S_{b}$}.\\
-K(S_{b}/S_{\rm ref})^{\gamma_{2}-\gamma_{1}}(S/S_{\rm ref})^{-\gamma_{2}},   & \text{$S > S_{b}$}
\end{cases}
\end{equation}

\begin{equation}
N(>S) =\begin{cases}
 K\Big (\frac{1}{1-\gamma_{1}}-\frac{1}{1-\gamma_{2}} \Big ) \Big(\frac{S_{b}}{S_{\rm ref}}\Big )^{(1-\gamma_{1})}\\
 +K\Big(\frac{1}{\gamma_{1}-1}\Big) \Big(\frac{S}{S_{\rm ref}} \Big )^{1-\gamma_{1}}, & \text{$S <S_{b}$}\\
-K\Big(\frac{1}{\gamma_{2}-1}\Big) \Big(\frac{S_{b}}{S_{\rm ref}}\Big)^{(\gamma{2}-\gamma_{1})}\Big(\frac{S}{S_{\rm ref}}\Big)^{1-\gamma_{2}},   & \text{$S > S_{b}$}
\end{cases}
\end{equation}
where the $\gamma_{1}$ and $\gamma_{2}$, are the faint- and bright- end slopes, K is a normalization factor, and $S_{b}$ is a break flux.   $S_{\rm ref}$ is a normalization flux. which we set to $10^{-15} \rm erg\ s^{-1} cm^{-2}$.

In our fitting, we adopt four flux ranges with different minimum fluxes, each corresponding to the 20\%, 50\%, 80\%, and 90\% completeness levels, which are presented in Table~3, in order 
to investigate the effect from sensitivity. The maximum flux density is the same {\bf for all four} subsamples ($3.74\times 10^{-13}$, $1.49\times 10^{-13}$, $2.16\times 10^{-13} \rm \ erg\ s^{-1} \ cm^{-2}$ for full-, soft-, and hard- bands respectively).
To estimate the uncertainties of the model parameters, we conducted Monte Carlo simulations 
with 1000 mock distributions generated by using errors calculated following \cite{Gehrels1986}.  We then refit the model to each mock distribution.
We take the median and the standard deviation of the N=1000 distributions for each parameter as the best-fit value and the error bar.
We note that we fit the differential and cumulative number count distributions separately, which yielded  different model parameters.
This might be due to the small number of detections in the differential number counts in the high flux regime, and hence  
uncertainties for the differential number counts are much larger than those for the cumulative number counts.
We provide the model parameters for the differential and cumulative number count distributions for each completeness level in each band in Table 4. 

As examples we present the differential and cumulative number count distribution in the hard band with the best-fit model (red solid) 
for the four various fitting ranges in Figures~7 and 8, respectively. The black symbols (in Figure~7) and lines  (in Figure~8) denote the data included in each fit.   
In each panel, we also plot distributions from the Chandra Multiwavelength Project (ChaMP) fields  \citep[purple dotted lines;][]{Kim2007}
and the CDF-S fields \citep[blue dashed lines;][]{Lehmer2012} for comparison.

As shown in Table 4 and in Figure~7 and 8, we find that the best-fit model parameters vary depending on the flux range included 
in the fit. It seems that including lower fluxes in the fit causes $\gamma_{1}$ to be higher and $\gamma_{2}$ to be lower. 
The normalization factor and the break flux also vary, but they do not show significant dependences on the adopted minimum flux.
As shown in Figure~7, the differential number counts below $\sim10^{-15} \rm erg\ s^{-1} cm^{-2}$ 
are smaller as flux decreases, and this feature affects the fitting results by including or excluding such low number counts
at the faint flux end. Similarly to the hard band, fitting results for distributions in full-, and soft- bands also show strong
dependency on fitting range (Table 4).
This result suggests that statistical uncertainties can be an issue in the low flux regime, hence
the fitting range range should be carefully considered in the analysis of number count distributions. 

Interestingly, there are large differences even between previous works except for $\gamma_{2}$ (consistent within uncertainties). For example, for the distribution in the hard band, $\gamma_{1}$ are 1.82$\pm$0.01 and 1.58$\pm$0.01 for ChaMP and ChaMP+CDF-S \citep{Kim2007} and 1.32$\pm$0.04 for CDF-S \citep{Lehmer2012} giving $\sim0.5$ dex difference.  This shows that the low flux regime is more sensitive to the adopted completeness level than the high flux regime and that one needs to consider the fitting range carefully.
\\

\begin{figure*}{}
\centering
\includegraphics[width = 0.94\textwidth]{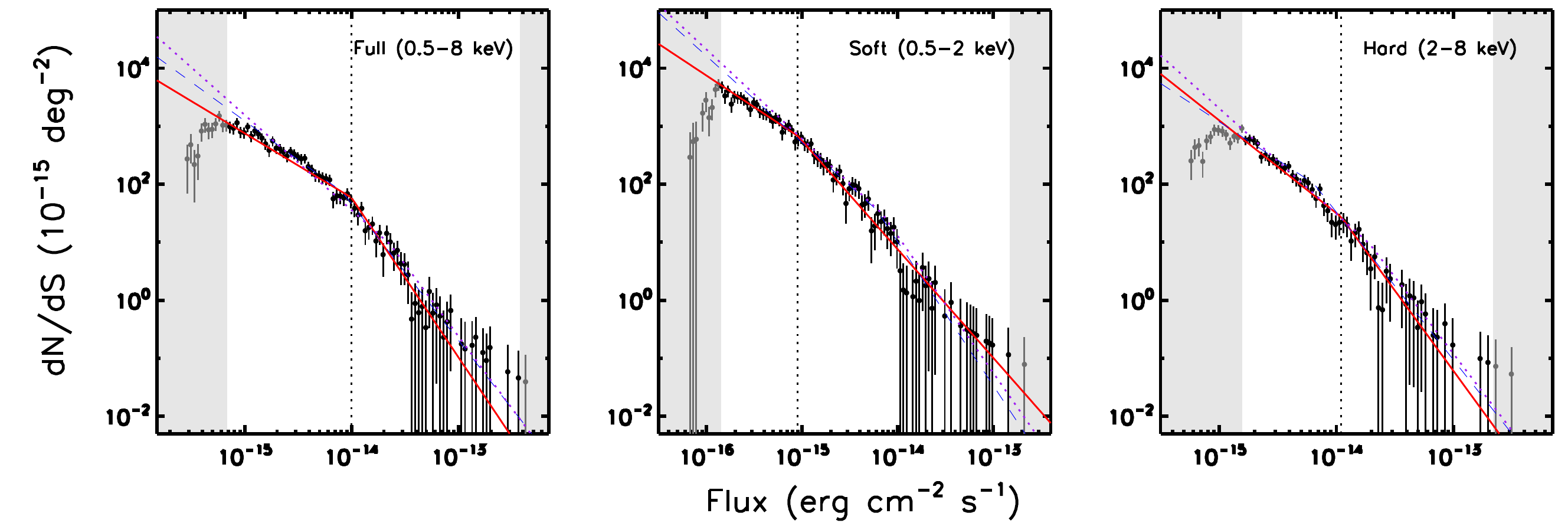}
\caption{
Differential number count distribution in the full-, soft-, and hard- bands with best-fit model (red) fitted over the 50\%\ sensitivity level. Symbols and colors are same with Figure~7.
\label{fig:allspec1}}
\end{figure*}

\begin{figure*}{}
\centering
\includegraphics[width = 0.94\textwidth]{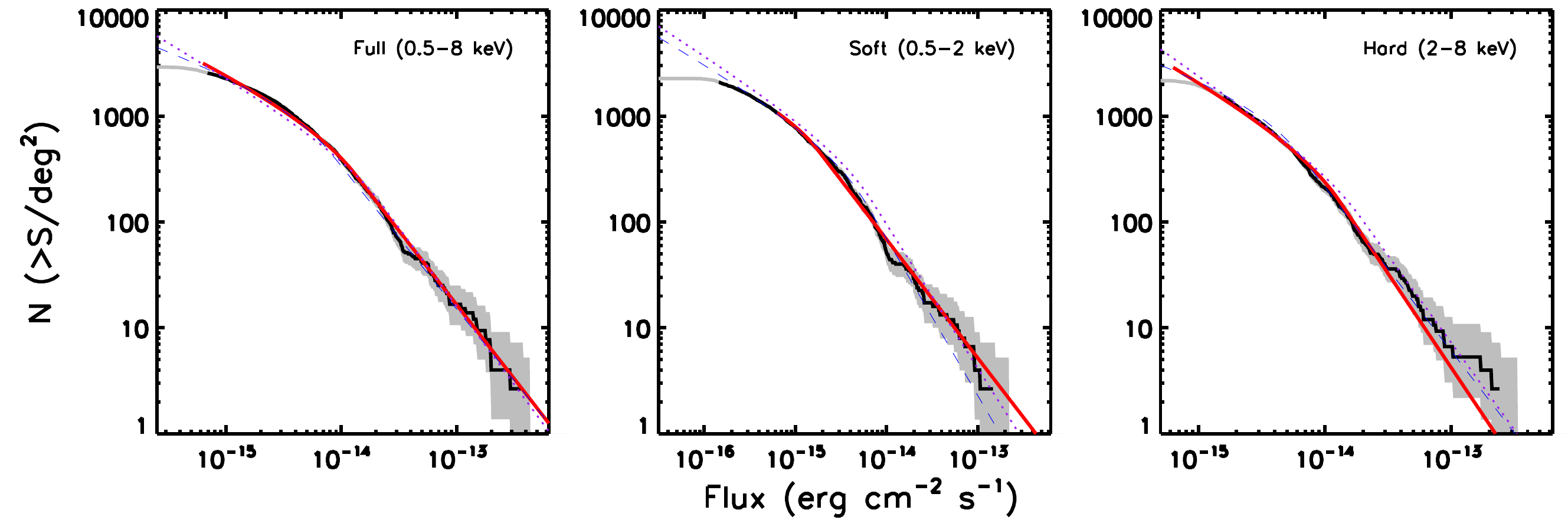}
\caption{
Cumulative number count distribution in the full-, soft-, and hard- bands with best-fit model (red) fitted over the 50\%\ sensitivity level. Symbols and colors are same with Figure~8.
\label{fig:allspec1}}
\end{figure*}

\subsection{Normalization in number count distribution}\label{astrometry correction}
In Figures~9 and 10, we plot the differential- and cumulative- number counts in the three bands with the best-fit model fitted over the 50\%\ sensitivity level. Interestingly, our results, at least in the soft- and hard- bands, show somewhat lower normalizations to the cumulative number count distribution compared to the distributions of X-ray point sources in the fields \citep{Kim2007,Gilmour2009,Lehmer2012,Ehlert2013}. For example, from 50\% completeness level to $10^{-13}\ {\rm erg\,s^{-1}\,cm^{-2}}$,
our distribution in soft- and hard- bands are on average $\sim22$\%\ lower than the that of the ChaMP. Below, we discuss possible reasons for this discrepancy, including 1) environmental effects, 2) cosmic variance, and 3) systematic errors.

First, the large-scale cluster environment can potentially affect the number of detected X-ray point sources \citep[e.g.,][]{Gilmour2009, Ehlert2013}. However, we do not expect environmental effects to be significantly altering the $\log N-\log S$ distribution across the Abell 133 field. For example, if AGN activities are suppressed in a cluster environment, then the $\log N-\log S$ distribution should be similar compared to that in a non-cluster environment, since the number of  foreground/background cosmic X-ray sources will be similar regardless of environmental effects. On the other hand, if AGN activities are enhanced within cluster environments \citep[e.g.,][]{Gilmour2009, Ehlert2013},  then the $\log N-\log S$ distribution should have a higher normalization. Thus, it is  difficult to understand how the enhanced AGN activities in a cluster environment can yield the \textit{lower} $\log N-\log S$ normalization that we  observe in the Abell 133 field. 

Second, we consider the possibility of a different normalization due to cosmic variance. \cite{Luo2008} reported that the normalization of the $\log N-\log S$ distribution from the CDF-S is $\sim$ 25\% smaller than that from the CDF-N, suggesting a possibility that the difference may be caused by field to field variations. Wide-field surveys of more regions of the sky are required to quantify the level to which the normalization of $\log N-\log S$ might vary from field to field.  

Finally, there could be systematic uncertainties introduced from different (or inconsistent) analyses (i.e., including lower fluxes in the fitting) among various works. To find out the systematic differences among various studies is beyond the scope of this study. 
\\

\subsection{Radial distribution of number densities}

The inner $\approx$3$\arcmin$ of the Abell 133 cluster contains diffuse X-ray emission from hot gas (see Figure~1), which could degrade the sensitivity for detecting point sources in the inner region compared to the cluster outskirts. Below, we investigate if the presence of diffuse emission can affect our $\log N - \log S$ distribution, particularly for the low-flux sources. 

We separate the X-ray point sources into two groups, which are divided by the flux sensitivity limit at the 80\% completeness. We then compare the source count distributions in the three bands as a function of the distance from the center of the  cluster, which is determined as the location of peak diffuse X-ray emission as RA: 15.6737$\arcdeg$, DEC: -21.8815$\arcdeg$ (see Figure~11).

The subset of the high-flux objects shows a relatively flat distribution with the radius, which is expected if the majority of the point sources are foreground/background objects. In contrast, we find a small peak at the innermost region ($<$0.25 Mpc, or 3.8 \arcmin), while the region beyond $R_{200}$ (i.e., $\gtrsim$1.6 Mpc or 24.2 \arcmin) shows a decline in the source counts, plausibly due to the poor sensitivity at large radial distance (see Figure~5). We note that the sensitivity inside of $R_{200}$ is mostly higher than the 80\% completeness, meaning that the high flux sample is not highly incomplete within $R_{200}$ because of low-sensitivity. The peak in the central region suggests that we are not missing high-flux sources embedded in the diffuse emission. If we assume that the contamination of the non-cluster member X-ray point sources is constant along the radius, the central peak in the radial distribution is consistent with the results reported in the previous studies \citep[e.g.,][]{Gilmour2009, Ehlert2013}. Since the galaxy number density is steeply decreasing as a function of radius, the weak increase of the X-ray point source at the center is still consistent with the radial decrease of AGN fraction. 
For example, the galaxy number density typically decreases by $\sim$an order of magnitude from R$_{500}$ to the center \citep[e.g.,][]{Popesso2007}, while the number density of the X-ray point source decreases by only a factor of a few in Abell 133. In Figure~12, we present the AGN fraction for the high flux subsamples, which is calculated as \cite[e.g.,][]{Ehlert2013}
\begin{equation}
\phi = \frac{N_{X}(r)-C_{x}}{N_{G}(r)},
\end{equation}
where $N_{X}(r)$ and ${N_{G}(r)}$ are the number densities of X-ray point sources and galaxies. The galaxy number density was estimated assuming a King model with a core radius of $r_{c}/r_{200}$=0.224 (see \citealt{Popesso2007}). $C_{x}$ is the number density of the  background, and we adopted the values of 646, 556, and 302 for full-, soft-, and hard- bands, respectively, which are the average of the number densities at 2 Mpc and 2.25 Mpc. The observed radial trend in Abell 133 would imply that AGN fraction decreases toward the cluster center, suggesting the lower level of AGN triggering in denser environments \citep[e.g.,][]{Gilmour2009, Ehlert2013}. We note that the AGN fraction calculated here is sensitive to the background, which was arbitrarily selected. The reason for the arbitrary selection is that we do not know the actual number density of the background. Thus, at <1.6 Mpc, the uncertain background subtraction may lead to the positive slope shown in Figure~12. For a more reliable estimation, individual cluster members should be identified, which is beyond the scope of this paper. We also note that the decrease in AGN fraction at >1.6 Mpc is likely caused by poor sensitivity.

On the other hand, the trend for the low-flux objects is different.  For example,  we detect the lower number of the low-flux sources in the innermost region ($<$0.25 Mpc), which is likely caused by the low-flux sources being diluted below the detection threshold by the diffuse X-ray emission. Interestingly, we find an excess of low-flux sources at moderate radii, i.e., at 1.0--1.5 Mpc. This trend could be due to a sensitivity effect. As seen in Figure 5, the sensitivity at 1.0--1.5 Mpc is higher than other regions, hence we may detect more low-flux sources. To understand this, we took average sensitivities in the hard band at 0.5--1.0 Mpc ($10^{-14.5}\, \rm erg\ s^{-1} cm^{-2}$) and 1.0--1.5 Mpc ($10^{-15.1}\, \rm erg\ s^{-1} cm^{-2}$) and compared the expected number densities based in each radial bin (using our cumulative number count distribution). The number densities are 1133 (at 0.5--1.0 Mpc) and 1956 (at 1.0--1.5 Mpc), consistent with our results (Figure~8). In addition to sensivity, gas clumpiness could be another factor influencing this trend. By investigating gas clumpiness in Abell 133 out to $R_{200}$, for example, \citet{Morandi2014} reported that the gas clumping factor at radii beyond 1 Mpc (which corresponds to $R_{500}$) is larger than unity (i.e., a factor of 2-3). They interpreted that this result indicates that the hot gas is virialized interior to $R_{500}$, but not at larger radii. While one effect of diffuse emission would be to lower the sensitivity for detecting point sources, it is plausible that the clumpy gas at the moderate radius is instead introducing an excess of spurious low-flux X-ray sources, where gas clumps are improperly identified as points sources. We stress that effects from diffuse gas emission are not of concern for our high-flux subset. In other words, the diffuse emission does not affect the normalization of our $\log N-\log S$ distribution.   \\

\begin{figure*}{}
\centering
\includegraphics[width = 0.94\textwidth]{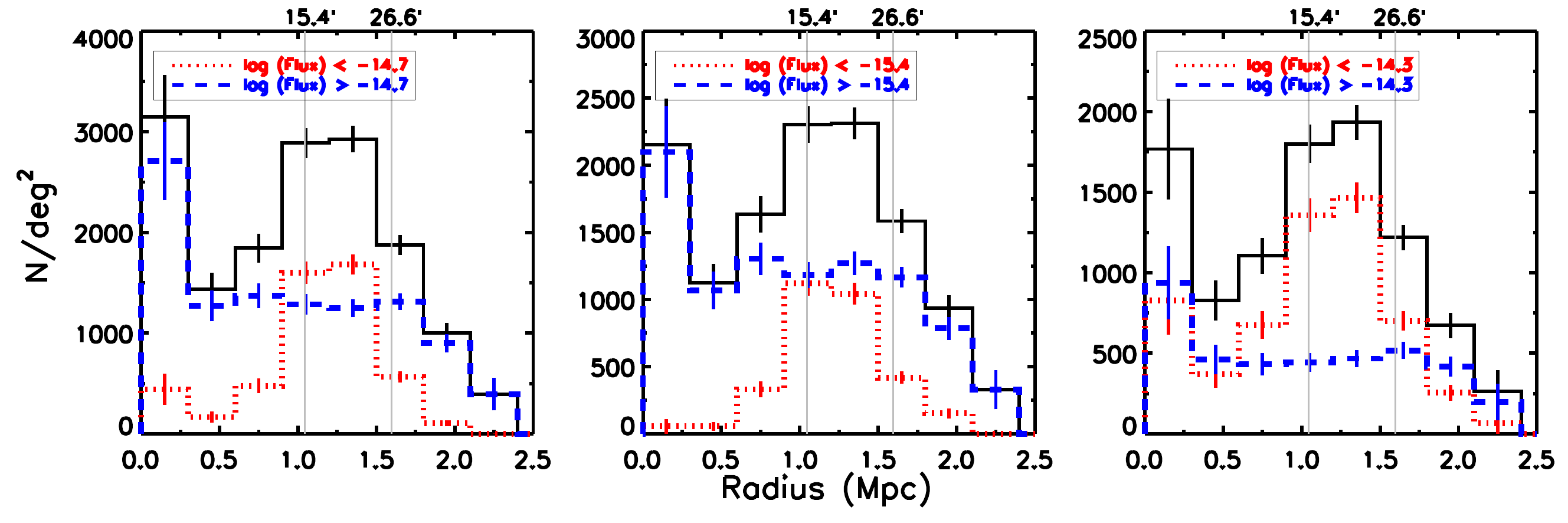}
\caption{
The number density of sources in the full- (left), soft- (center), and hard- (right) bands as a function of distance from the cluster center.  Black, red, and blue lines represent the full sample, the low flux subsample ($\log f_{\rm 0.5-8 keV} < -14.7$, $\log f_{\rm 0.5-2 keV} < -15.4$, and $\log f_{\rm 2-8 keV} < -14.3$), the and
high flux subsample ($\log f_{\rm 0.5-8 keV} > -14.7$, $\log f_{\rm 0.5-2 keV} > -15.4$, and $\log f_{\rm 2-8 keV} > -14.3$), respectively. $R_{500}$ (left) and $R_{200}$ (right) are denoted as vertical gray lines. The uncertainties from Poisson statistic are shown at each radius bin.
\label{fig:allspec1}}
\end{figure*} 

\begin{figure*}{}
\centering
\includegraphics[width = 0.94\textwidth]{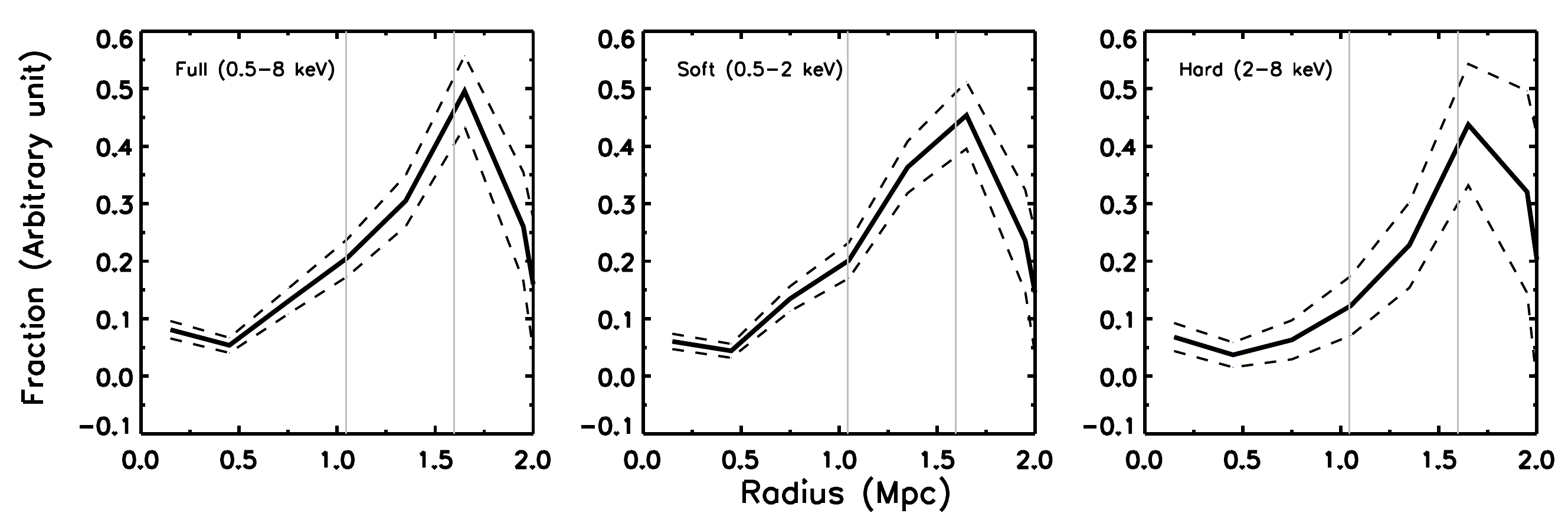}
\caption{
X-ray point source fraction as a function of distance from the cluster center for high flux subsample ($\log f_{\rm 0.5-8 keV} > -14.7$, $\log f_{\rm 0.5-2 keV} > -15.4$, and $\log f_{\rm 2-8 keV} > -14.3$). The dashed line shows the uncertainties from Poisson statistic. 
$R_{500}$ (left) and $R_{200}$ (right) are denoted as vertical gray lines.
\label{fig:allspec1}}
\end{figure*} 

\section{Summary}\label{summary}
We investigate X-ray point sources in a wide-field \textit{Chandra} image (0.76 deg$^2$) covering the Abell 133 cluster of galaxies at $z=0.0566$. From the relative deep X-ray mosaic images with a total exposure time of 2.8 Msec, we detect 1617, 1324, and 1028 X-ray point sources in the full-, soft-, and hard- bands with the flux limits at the 50\%\ completeness level of $6.95 \times 10^{-16}, 1.43 \times 10^{-16}, \rm \ and \ 1.57 \times 10^{-15}$ \ergs\ $\rm cm^{-2}$.. We present the X-ray point source catalog, which can be utilized for future multiwavelength studies of Abell 133 and its environment. The combined number counts of the X-ray point sources across the entire field are well represented with a broken power-law and the fitting parameters are in agreement with those of X-ray point source distributions in other fields. Similar to previous works \citep[e.g.,][]{Gilmour2009,Ehlert2013}, we find a slight excess of X-ray point sources that could be associated with the cluster, suggesting that AGN fraction decreases toward the cluster center, for given an increase of galaxy number density at the center. This result is consistent with the suppressed AGN activities in dense environments. 
  \\

\acknowledgements
This work has been supported by the Basic Science Research Program through 
the National Research Foundation of Korea government (2016R1A2B3011457 and No.2017R1A5A1070354).
This paper used data taken from the CFHT Science Archive. 
This research has made use of data obtained from the Chandra Data Archive and software provided by the 
Chandra X-ray Center (CXC) in the application packages CIAO. 

\bibliography{Abell133}

\end{document}